\documentclass[11pt,aps,showpacs,preprintnumbers,amsmath,amssymb]{revtex4}
\usepackage{amsmath,amssymb,graphicx}
\usepackage{epsf}
\usepackage {amssymb}
\usepackage{upquote}
\usepackage{multirow}
\usepackage{color}
\newcommand{\nc}{\newcommand}
\nc{\ba}{\begin{eqnarray}} \nc{\ea}{\end{eqnarray}}
\nc{\be}{\begin{equation}} \nc{\ee}{\end{equation}}

\newcommand\s{\sigma}

\newcommand\e{\epsilon}

\newcommand\la{\lambda}

\nc{\ga}{\gamma} \nc{\x}{{\bf x }} \nc{\kk}{{\bf k }} \nc{\f}{{\bf f
}} \nc{\T}{ \theta (s_i (t)- \s) } \nc{\TT}{ \theta (s_i (t_{ r \, i
} )- \s) } \nc{\br}{   (s_i (t)- \s)  } \nc{\fa}{\phi_1}
\nc{\fb}{\phi_2}

\input{epsf.sty}
\begin{document}

\title{ String melting in a photon bath}
\preprint{MIT-CTP 4422}
\author{Johanna Karouby\footnote{karoubyj@mit.edu}}
\affiliation{Center for Theoretical Physics and Department of Physics, \\ Massachusetts Institute of Technology, Cambridge, Massachussetts 02139, USA}

\begin{abstract}
 We compute the decay rate of a metastable cosmic string in contact with a thermal bath by finding the instanton solution.
 The new feature is that this decay rate is found in the context of non thermal scalar fields in contact with a thermal bath of photons.
In general, to make topologically unstable strings stable, one can couple them to such a bath. The resulting plasma effect creates metastable configurations which can decay from the false vacuum to the true vacuum. In our specific set-up, the instanton computation is realized for the case of two out-of-equilibrium complex scalar fields : one is charged and coupled to the photon field, and the other is neutral.
New effects coming from the thermal bath of photons make the radius of the nucleated bubble and most of the relevant physical quantities temperature-dependent.
However, the temperature appears in a different way than in the purely thermal case, where all scalar fields are in thermal equilibrium.
As a result of the tunneling, the core of the initial string melts while bubbles of true vacuum expand at the speed of light. 
 \end{abstract}

\maketitle

\section{Introduction}

Cosmic strings are linear topological defects arising when the first homotopy group of a vacuum manifold is non trivial.
If stable, topological defects can play an important role in the early universe (see e.g. \cite{ShellVil, HK, RHBrev} for overviews).
More specifically cosmic strings can contribute to structure formation, play a role in baryogenesis \cite{Brandenberger:1994bx},  explain the origin and coherence of cosmological magnetic fields on galactic scale \cite{CSmag} and even
modify temperature fluctuations and non-gaussianities of the CMB \cite{Tashiro:2012pp}\cite{Bevis:2007gh}. They also form loops that can contribute to ultra-high-energy cosmic rays \cite{Bhattacharjee:1998qc} and arise in some string theory models after brane inflation.

In the following, we focus on the decay of field-theoretic superconducting strings coupled to a thermal bath of photons.
The key idea is to compute the decay rate for metastable strings previously stabilized by a photon plasma.
Here we work with an embedded string made of two non-thermal complex scalar fields, where one is neutral and one is charged.
Since charged fields are naturally coupled to photon gauge fields, a thermal bath of photons can easily be implemented in such cases.
We assume a temperature below a certain critical temperature where the scalar fields are still out of equilibrium. This assumption is important since the resulting effective potential does not contain any thermal terms for the scalar fields.

In our approach we focus on first order phase transitions.
Such transitions usually imply a period of non-equilibrium physics and typically occur via bubble nucleation. As a result, cosmological relics such as topological defects can form.
In the following we study the melting of a superconducting
string in models analogous to the Ginzburg-Landau theory of superconductivity \cite{Nielsen:1973cs}.
Our superconducting string is made of a neutral complex field $\phi$ and an initially vanishing charged complex field $\pi_c$.

As a toy model for analytical study of the stabilization of
embedded defects by a plasma, we start by considering the chiral limit of the
QCD linear sigma model. The Lagrangian describing the system involving the sigma field $\sigma$ and the
 pion triplet ${\vec \pi} = (\pi^0, \pi^1, \pi^2)$ is
\begin{equation}
\label{lag2a}
{\cal L} \, = \, {1 \over 2} \partial_{\mu} \sigma \partial^{\mu}
\sigma + {1 \over 2} \partial_{\mu} \pi^0 \partial^{\mu} \pi^0 +
D_{\mu}^+ \pi^+ D^{\mu -} \pi^- -  {\lambda \over 4} (\sigma^2 + {\vec \pi}^2 - \eta^2)^2,
\end{equation}
where
$D_{\mu}^+ \, = \, \partial_{\mu} + i e A_{\mu} \, , \,\,\,\, D_{\mu}^-
\, = \, \partial_{\mu} - i e A_{\mu} $, and $\eta^2$ is the ground state expectation value of $\sigma^2 + {\vec
\pi}^2$.
For simplicity we use two complex scalar fields: one combining the charged pions $\pi_c = \pi^1 +i\pi^2 $ and the other
representing the neutral pions $\phi = \s+i \pi_0$.
 In the core of the string $\phi=0$ so that
we are effectively left with a complex scalar field $\pi_c$ and a U(1) gauge symmetry. As proved in \cite{Halperin:1973jh}, this configuration admits a first order phase transition with a metastable potential.
As a result the system can move to its lowest energy state, the true vacuum, through a process known as bubble nucleation. The goal of this paper is to quantify this quantum tunneling from the false to the true vacuum by finding the instanton solution.

Our approach to instanton computation closely resembles the study of first order phase transitions in the early universe \cite{Quiros:1999jp,Kibble:1976sj,Linde:1981zj,Dolan:1973qd,Dine:1992wr}. The difference here is that we study the breaking of strings in contact with a thermal bath. This string breaking through instanton production has already been studied in the literature, but not in presence of a thermal bath\cite{Holman:1992rv,Dasgupta:1997kn,Garriga:1991tb,Preskill:1992ck, Monin:2008mp, Monin:2008uj}.
It is important to remember that the thermal bath we consider is in contact with an out-of-equilibrium string: The resulting effective potential has been found in \cite{Karouby:2012yz} and will be the starting point of our analysis.

As a concrete example, we first introduce the linear sigma model, which describes pions at low energy.
In this set-up, the pion string is topologically unstable due to the fact that the vacuum manifold symmetry group O(4) has a trivial first homotopy group.
In a previous attempt to make the pion string stable, we found that the presence of a photon bath enhances the stability of the string for a certain range of temperatures by effectively reducing the vacuum manifold to a circle  \cite{Nagasawa:1999iv}.
However, the string is not stable but rather metastable, provided we choose the value of the quartic coupling accurately \cite{Karouby:2012yz}.
This metastability implies the production of an instanton when the fields tunnel to a lower energy state.

Second, we review an example of a first order phase transition at finite temperature.
 Due to the shape of the potential, the string is stable above a certain temperature, metastable at smaller temperatures and classically
 unstable below some temperature $T_0$.

In the appendix we introduce an ansatz for the string breaking at zero temperature. Each pair of fields describing the string has an initial and final configuration, before ($t=-\infty$) and after ($t=0$) tunneling. In the case at hand, the string breaks at the point where it nucleates a cylindrical instanton, which expands along the string at almost the speed of light. In the case where monopole-anti-monopole pairs are nucleated, we expect the speed of this expansion to be reduced.

Finally, in order to analytically estimate the decay rate of the string into two strings we consider spherical instantons .
 We focus on the semi-classical bounce computation by Coleman, which
considers a thin wall \cite{Coleman:1977py}.
The quantum mechanical decay at zero temperature for various topological defects in 1,2 or 3 spatial dimensions has been discussed fairly comprehensively in \cite{Preskill:1992ck}:  for example, a string breaking due to monopole-anti monopole nucleation yields the decay rate  $ \Gamma  \sim e^{- \frac{\pi m^2}{\mu}}$ where $m$ is the monopole mass and $\mu$ the string tension.
Here we extend the computation to embedded defects in contact with a plasma.
We compute the decay rate of the string (or bubble nucleation rate) $ \Gamma  \sim e^{-B}$ where B is the bounce action in euclidean space.
To do so we use the thin-wall approximation which allows us to estimate the euclidean action for the spherical instanton.
 As a result, we find the temperature-dependent radius of the nucleated bubble, the decay rate for precise values of the parameters, and conclude on how likely  the melting process is.

 The string will be truly stable and thus have an impact on cosmology, only if the string decay rate is smaller than the Hubble parameter H :
  $ \Gamma < H$. If the decay rate is large enough, the core of the string melts and the string breaks.

\section{Effective potential for the neutral string}

\subsection{The pion string}

Here we simply recall results from \cite{Karouby:2012yz}.
We only consider out-of-equilibrium scalar fields so the usual assumption that topological defects are out of equilibrium still holds.
Only the gauge field describing the photon plasma in contact with the string is thermal.
For simplicity we use units where $\hbar=c=1$ and do not consider the expansion of the universe for now.

As a toy model for analytical study of the stabilization of
embedded defects through plasma effects, we start by considering the chiral limit of the
QCD linear sigma model. The Lagrangian describing the system involving the sigma field $\sigma$ and the
 pion triplet ${\vec \pi} = (\pi^0, \pi^1, \pi^2)$ is given by
\begin{equation}
\label{lag1}
{\cal L}_0 \, = \, {1 \over 2} \partial_{\mu} \sigma \partial^{\mu}
\sigma + {1 \over 2} \partial_{\mu} {\vec \pi} \partial^{\mu} {\vec
\pi} - {\lambda \over 4} (\sigma^2 + {\vec \pi}^2 - \eta^2)^2 \, ,
\end{equation}
where $\eta^2$ is the ground state expectation value of $\sigma^2 + {\vec
\pi}^2$.

Two of the scalar fields, the $\sigma$ and $\pi_0$, are electrically
neutral, while the other two are charged.
The above Lagrangian is invariant under O(4) rotations, including the global rephasing (or global U(1) symmetry) of the charged fields:
 $(\pi^1 +i\pi^2) \rightarrow e^{i\omega}(\pi^1 +i\pi^2)$ where
$\omega$ is a constant phase.\\
Gauging this symmetry by coupling the charged fields to the photon field gives
\begin{equation}
\label{lag2}
{\cal L} \, = \, {1 \over 2} \partial_{\mu} \sigma \partial^{\mu}
\sigma + {1 \over 2} \partial_{\mu} \pi^0 \partial^{\mu} \pi^0 +
D_{\mu}^+ \pi^+ D^{\mu -} \pi^- - V_0 \, ,
\end{equation}
where
$D_{\mu}^+ \, = \, \partial_{\mu} + i e A_{\mu} \, , \,\,\,\, D_{\mu}^-
\, = \, \partial_{\mu} - i e A_{\mu} \, $,
$\pi^+ \, = {1 \over {\sqrt{2}}} (\pi^1 + i \pi^2)$ ,
$\pi^- \, = {1 \over {\sqrt{2}}} (\pi^1 - i \pi^2) $
 and $V_0$ is a Mexican-hat potential, $V_0  =  {\lambda \over 4} (\sigma^2 + {\vec \pi}^2 - \eta^2)^2 $.

The result of gauging the symmetry is to add a new effective pion-photon interactions through the covariant derivative in (\ref{lag2}).
This new interaction with the photon field allows the subsequent implementation of a thermal bath of photons.
The presence of this thermal bath reduces the vacuum manifold from a 3-sphere to a circle: ${\cal{M}}_i=S^3 \rightarrow {\cal M}_f=S^1$.
More precisely the initial vacuum manifold ${\cal{M}}_i$ is a 3-sphere of radius $\eta$, such that  $$(\pi^1)^2+ (\pi^2) \,^2+\sigma^2 +{\pi_0}^2 =\eta^2$$
The thermal bath reduces this sphere to a circle of same radius, ${\cal{M}}_f$ : $$\sigma^2 +{\pi_0}^2 =\eta^2 \ \ and  \ \ \pi^+ = \pi^-=0$$

Before adding a thermal bath, string configurations exist but are not topologically stable since the first homotopy group
 of the 3-sphere is trivial.
 However other topological defects like skyrmions (or texture) can form since $\pi_3(S^3)\neq1$.
 Here the role of the thermal bath of photons is to modify the effective potential and stabilize the strings topologically since, after adding a plasma, the vacuum manifold
has a non trivial first homotopy group, $\pi_1({\cal{M}}_f=S^1)\neq1$.\\

\subsection{Pion string stability and effective potential}

In the linear sigma model of QCD, one has for realistic values of $\la_{QCD}$, $\frac{\la}{e^2}  \sim O (10^2) $.
However in a previous attempt to make the embedded string stable by using plasma effects, the condition for stability found is $\frac{\la}{e^2} \gg 1 $\cite{Nagasawa:1999iv}.
This implies that the pion string as it is described cannot be stabilized by plasma effects for realistic values of $\la_{QCD}$.
However, the effective potential found in \cite{Karouby:2012yz} is slightly different and metastable
\ba
\label{V1}
V_{eff} (\phi,\pi_c,T)\hspace{-1mm} =\hspace{-1mm}V_0\hspace{-1mm}-\hspace{-1mm}\frac{\pi^2 T^4}{45}
\hspace{-1mm}+\hspace{-1mm} \frac{e^2|\pi_c|^2 T^2}{12}\hspace{-1mm}-\hspace{-1mm}\frac{e^3|\pi_c|^3  T}{6\pi} - \frac{e^4 |\pi_c|^4}{16\pi^2}\hspace{-1.5mm} \left[\ln\hspace{-1.5mm}\left(\frac{e |\pi_c| e^{\gamma_{E}}}{4\pi T}\right)\hspace{-1.6mm}-\hspace{-1mm}\frac{3}{4}\right]\hspace{-1.5mm}+\hspace{-1mm}{\cal{O}}(\frac{e^6 \pi_c^6}{T^2})
\ea
where $\gamma_{E}$ is the Euler-Mascheroni constant.\\

For this potential, the vacuum manifold forms a circle in phase space $|\phi|^2=\eta^2  \ , \ \ |\pi_c|=0$ and so the system admits string solutions.
In this case, at temperature below the confinement scale, an embedded neutral string exists.

Since the above potential is a one-loop effective potential it is only valid if the expansion parameter $x$ in the perturbative expansion is small enough
\ba
\label{ex}
x=\frac{\la T}{m_{eff}} \ll 1
 \ea
where $m_{eff} = e \pi_c$ is the effective mass.
In order to estimate the expansion parameter in our model, we simply equalize the cubic and quartic terms in  (\ref{V1})to find when the broken minimum appears
\ba
\la \pi_c \sim e^3 T
\ea
which in turn implies that $ \pi_c \sim \frac{e^3 T}{ \la}$.
Plugging back into (\ref{ex}) yields $$x=\frac{\la^2}{e^4} $$
For the perturbative expansion to be valid we thus require   \ba \frac{\la^2}{e^4}  \ll 1 \label{co1} \ea
Unfortunately, for realistic values of $\la_{QCD}$ the above condition is not satisfied.
However, what we have found is consistent with the fact that first order effective potentials in QCD cannot be trusted since higher loop corrections are large close to the degenerate minimum \cite{Arnold:1992rz}.
In the following we therefore choose to work in a regime where the perturbative expansion is valid and so we can trust the one-loop effective potential
(\ref{V1}).

\subsection{The superconducting string}
From now on we work in the regime where the effective potential (\ref{V1}) is valid and the 2 complex charged and neutral scalar fields, $\pi_c $ and
$\phi $,  do not represent pions anymore.
The crucial point is that there are four fields so the tunneling process is, in general, extremely difficult to estimate.
In order to simplify the study of the decay of the neutral string, we assume tunneling occurs in the core of the string so that $\phi \sim 0$ during tunneling.
As a result, in its core, the neutral string sees a potential barrier only in the charged field direction.

In contrast to what happens in the core of the string, outside the core the total configuration is already in its true vacuum and cannot decay to a lower energy state.
 The energy level is already zero so the state cannot decay to a lower energy state: no tunneling occurs.
It is thus well justified to consider quantum tunneling only in the core of the string.

\begin{figure}[htbp]
\centering
\includegraphics[scale=0.20]{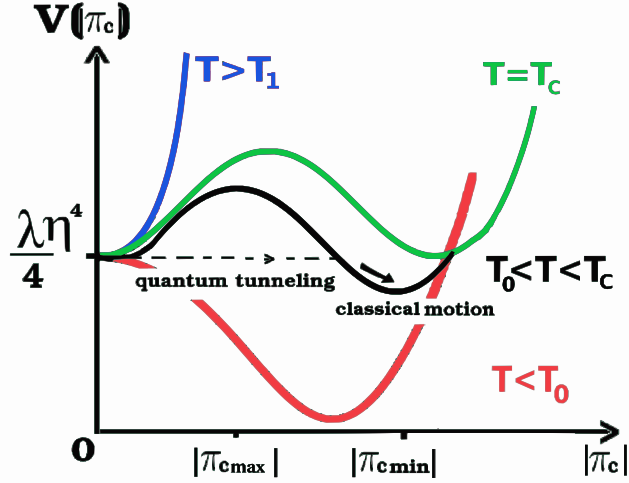}
\caption{Finite temperature effective potential in the core of the string.
The horizontal dashed line indicates the tunneling of the string from its initial configuration (where the charged fields vanish
everywhere) to the exit point after quantum tunneling.
The arrow toward the minimum indicates the classical decay towards the true vacuum which occurs after quantum tunneling. }
\label{tun}
\end{figure}

 Considering tunneling only in the core
amounts to setting the neutral field $\phi$ to zero during the whole process, so the decay to the false vacuum only occurs along the $\pi_c$ direction.
This greatly simplifies our task from a computational point of view since, as show in Fig.\ref{tun}, the potential becomes a function of $\pi_c$ only.

Finally we are left with a complex scalar field $\pi_c$, and a U(1) gauge symmetry.
These models are analogous to the Ginzburg-Landau theory of superconductivity which, as proved in \cite{Halperin:1973jh}, admits a first order phase transition when $\frac{\la}{e^2} \ll 1 $.
The resulting effective potential in the core of the string in the high-temperature limit, $\frac{|\pi_c|}{T} \ll 1$ is
\be
V_{eff}(\pi_c,T) \simeq \frac{\la}{4}(|\pi_c|^2-\eta^2)^2+\frac{e^2|\pi_c|^2 }{12}T^2-\frac{e^3|\pi_c|^3 }{6\pi}T
\label{veff}
\ee
where we have neglected higher-order terms of the full effective potential (\ref{V1}) and dropped the logarithmic part.
\hspace{-0.5cm}

 As shown in Fig.$\ref{tun}$, above a temperature $T_1$,the potential barrier is infinite in the charged field direction and no tunneling occurs.
For temperatures between the critical temperature, $T_c$ and $T_1$,  $|\pi_c|=0$ is the true vacuum so the string is topologically stable.
For both these cases the neutral string is also quantum mechanically stabilized.

However, for  temperatures between $T_0 $ and the critical temperature $T_c$,
the system exhibits a first order phase transition due to the negative cubic term in the potential.
As a result, the zero charged field condensate becomes metastable:  $\pi_c$ is in a  false vacuum so we expect quantum tunneling.
The charged fields tunnel to a non-vanishing expectation value in the true ground state, $\langle \pi_c \rangle=v'$ where $v'$ is
a complex number. It is important to note that non-trivial field configurations can occur since the field $\pi_c$ is complex and can take on any value
on the circle of radius $\lvert v' \rvert$.
 After tunneling, bubble instantons form and propagate along the string. As a result, the core of the neutral string melts and the string breaks at various locations.

In the following we compute $T_1$, the temperature above which the configuration is truly stable, and the critical temperature $T_c$, below which the string become metastable. We also compute all the particular field values necessary to study the instanton.

 For simplicity, we first rewrite the potential (\ref{veff}) in terms of powers of the charged field
\begin{equation}
V(\pi_c,T)=D(T^2-T_o^2)\pi_c^2-ET\pi_c^3+\frac{\lambda}{4}\pi_c^4
\label{V2}
\end{equation}
where the coefficients are given by
\begin{equation}
\label{valor}
D=\frac{e^2}{12} \ \ \
E=\frac{e^3}{6 \pi} \ \ \
T_0^2=\frac{6\la \eta^2}{e^2}
\end{equation}
and where the initial potential energy is set to 0 instead of $\frac{\la \eta^4}{4}$.

From (\ref{V2})it is easy to see that for temperatures below $T_c$ and above $T_0$ quantum tunneling occurs.
In non-expanding space, this tunneling occurs at constant energy and the charge field emerges at an exit point, $\pi_{c}^*$.

For temperatures below $T_0$, the fields classically roll down the potential and no tunneling occurs.
The string is still topologically stable. After rolling down the charge field would settle in the only minimum of the theory.
However, at temperatures close to $T_0$, the high temperature expansion becomes less and less accurate and
one must consider the initial effective potential (\ref{V1}) which includes higher powers of $\pi_c$. For this initial potential
the charged field stays frozen at $\pi_c=0$ and no classical rolling down occurs.

In order to find the values corresponding to the critical temperature $T_c$, the true vacuum $\pi_{c}-$ and the exit point $\pi_{c}^*$, we
first find the extrema of the potential.
This potential admits up to two extrema besides the false vacuum (which corresponds to $\pi_{c}=0$)
\be
\pi_{c}\mp=\frac{1}{2}(\frac{3ET}{\lambda}\pm\sqrt{9\frac{E^{2}T^{2}}{\lambda^{2}}-\frac{8D}{\lambda}\left(T^{2}-T_{0}^{2}\right)})
\label{exit}
\ee
The first solution, $\pi_{c}+$, is a local maximum and the second one, $\pi_{c}-$ is a local minimum. $\pi_{c}-$ also corresponds to the true vacuum $\pi_{c}-=v$ for temperatures
between $T_0$ and $T_c$.
These two solutions $\pi_{c}+$ and $\pi_{c}-$ only exist for a certain range of temperatures such that
\ba
 \frac{9E^{2}T^{2}}{\lambda}>8D\left(T^{2}-T_{0}^{2}\right)
\ea
Equating the two sides of this inequality defines a new threshold temperature $T_1$ above which
no tunneling occurs
\ba
T_{1}^{2}=\frac{8D\lambda T_{0}^{2}}{8D\lambda-9E^{2}}=\frac{6\lambda^{2}\eta^{2}}{e^{2}\lambda-\frac{6}{\pi^{2}}e^{6}}
\label{T1}
\ea
Besides since $T_1^2$ is positive the parameter $\la$ satisfies a new inequality
\ba \la \geq \frac{3 e^4}{8\pi^2}
\ea
Combining this inequality with the condition found in (\ref{co1}), constraints $\la$ to lie on a certain range of value in order to get tunneling
 \ba \frac{3 e^4}{8\pi^2} \leq \la \ll e^2 \label{lam1}\ea
Two other important quantities, the exit point at which the field emerges after tunneling, $\pi_c^*$ and the critical temperature,$T_c$ below which
quantum tunneling becomes possible, can be computed from the potential (\ref{V2}).
First, we find the zeros of the potential to find the exit point
\be
\label{p1}
\pi_{c}^*=\frac{2ET}{\lambda}\left(1-\sqrt{1-\frac{\lambda D}{E^{2}}\left(1-\frac{T_{0}^{2}}{T^{2}}\right)}\right)
\ee
Second, we find the temperature below which the potential possesses three zeros. This temperature corresponds to the critical temperature $T_c$
below which tunneling to the false vacuum starts. In order to find when tunneling start we simply require that an exit point exists.
The existence of an exit point forces the argument inside the square root of (\ref{p1}) to be positive and imposes an upper bound on the temperature
\be
T^{2}\geq\frac{\lambda D}{E^{2}}\left(T^{2}-T_{0}^{2}\right)\Rightarrow T^{2}\leq\frac{\lambda DT_{0}^{2}}{-E^{2}+\lambda D}=T_{c}^{2}
\ee
The above inequality defines the critical temperature, which can also be written in terms of the coupling constants $\la$ and the charge of the electron $e$
 \ba T_c=\sqrt{\frac{T_0^2}{1-\frac{e^4}{3\la \pi^2}} \label{TC}}.\ea
 As expected, this critical temperature is smaller than $T_1$ above which no tunneling occurs.
 We thus lowered our bound on the temperatures for which quantum tunneling is possible and for any temperature above $T_c$, the initial single string configuration remains stable. Below $T_c$ the string becomes metastable and quantum tunneling becomes allowed. It is important to remember that in any case, the neutral string is still topologically stabilized by the thermal bath of photons.\\

At $T=T_{c}$, the potential has two degenerate vacua located at $\pi_{c}=0$ and $\pi_{c}=\pi_c^D$. Moreover, since the two vacua have the same zero energy and that tunneling occurs at constant energy, the exit point after tunneling, $\pi_{c}^*$, is also the degenerate vacuum
\be
\;\;\pi_{c}^*(T_c)=\pi_c^D(T_c)=\frac{2ET_c}{\lambda}
\label{pD}
\ee

In case the two vacua are almost degenerate, one can use the thin-wall approximation to estimate the bubble nucleation rate through instanton production.
The instanton solution could be defined as the solution of the equation of motion interpolating between the initial configuration made up of one string
and the final one made up of two (see Appendix \ref{ap}. for a further description).
However here we restrict ourselves to the melting of the core of the neutral string and we just consider the tunneling process in the core of the string.
Therefore the charged field inside the string does not see the string background : in the core of the string, the neutral field $\phi$ vanishes and $\pi_c$ simply sees a potential barrier described by the metastable potential (\ref{V2}).
This greatly simplifies our computational task since we do not need to solve the equation of motion in the complicated background of the string.
The background becomes trivial and our problem now reduces to the study of the standard instanton with one scalar field tunneling from one false vacuum to the true vacuum.

\section{Spherical Instanton in a thermal bath of photons}

In the following we restrict ourselves to the study of the spherical instanton in the core of the string. This instanton describes the melting of the core of the neutral string due to tunneling
in the charged field direction. We assume the following :
\textit{\\
(i) $\phi \sim 0$ during the whole tunneling process. \\
(ii)The instanton nucleated is spherical in the 4 dimensional euclidean spacetime.  \\
(iii)The effective potential inside the core of the string is given by (\ref{V2}).}\\
(i) is justified by the fact that we only consider the core of the string. It also greatly simplifies the analytical estimate of the instanton action.\\
(ii) is justified since, amongst all the possible instanton configurations, the spherical instanton gives the smallest action and thus heavily dominates the decay rate.\\
(iii) comes from our particular model originally inspired from the pion string. The string configuration contains four non-thermal scalar fields in contact with a thermal bath of photons, and the effective potential
describing this model was derived in a previous work\cite{Karouby:2012yz}.

\subsection{Thin-wall approximation}

In order to estimate the value of the bubble nucleation rate we use the semi-classical approximation in the path integral formalism \cite{Coleman:1977py}.
The tunneling from the false vacuum $\langle \pi_c \rangle=0$ to $ \langle \pi_c \rangle \neq 0 $ is then described by the instanton solution which is a solution of the euclidean equation of motion.
Finding the euclidean action of the instanton $S_E$ and using a saddle point approximation gives the semi-classical decay rate :
\ba
\Gamma \propto e^{-S_E}
\ea
When the potential the barrier for the potential is thin  and the height of the barrier is large compared to the difference between the minima one can use the thin-wall approximation.
The thin-wall approximation uses the semi classical approximation and is valid when the bubble wall is thin.
In general, this occurs when the potential is almost degenerate so that the difference of energy between the true and false vacuum is always much smaller than the height
of the potential barrier.
For our potential (\ref{V2}), the approximation is valid for temperatures just below the critical temperature $T_c$.
Therefore the first instanton formed corresponds to thin-wall bubbles.
At lower temperature, the thin-wall approximation breaks and the wall thickens since the height of the barrier is small compared to the value of the potential at its minimum.
Whether bubble nucleation occurs via thick or thin-wall tunneling depends on
how big the decay rate is in the thin-wall case. If the decay rate is large enough, nucleation will proceed via thin-wall tunneling.
If not, one can solve the thick-wall case which requires numerical analysis.

Let us now  estimate the bubble nucleation rate in the thin-wall approximation.
We assume spherically symmetric solutions so that the equation of motion for the charged field has only one variable, the radial distance in spherical coordinates $r$
\ba
\frac{d^{2}\pi_{c}}{dr^{2}}+\frac{3}{r}\frac{d\pi_{c}}{dr}=V'(\pi_{c})
\ea
We consider the system at temperatures close to the critical temperature so that the difference in energy between the two minima is negligible.
 Since the two minima are almost degenerate, one can neglect the damping term in the equation of motion for the charged fields $\pi_c$.
 One can see this assumption makes sense by using a mechanical analogy \cite{Coleman:1977py}:

  Let us consider an upside down potential with a particle running from one maximum to the other. In this analogy, the first maximum corresponds to the true vacuum and the second one to the false vacuum. The scalar field evolution is described by the motion of the particle.
 Stating that the particle starts at one maximum and reaches the other one at infinite time, corresponds to stating that the scalar field starts in the
 false vacuum and reach the true one. So this scalar field does describe the instanton solution.
 Mechanically, for the particle to go from one maximum to the other one which has almost the same energy, the friction must be negligible and the energy $E_0=\frac{1}{2}\phi^ {\prime2}+V(\phi) $ is almost conserved.
In order not to lose too much energy the particle is initially close to the first maximum. Then it stays close to this maximum until some very large time $r \sim R$. Near time $R$ the particle moves quickly towards the valley and slowly comes at rest at the second maximum.
For the scalar field, this $R$ can be seen as the typical transition time at which the fields move to the true vacuum.
In euclidean space, R also corresponds to the radius of the bubble nucleated.

Neglecting the damping in term in the equation of motion for the charged field yields
 \ba
\frac{d^{2}\pi_{c}}{dr^{2}}=V'(\pi_{c})\simeq V_D'(\pi_{c})
\label{eq1}
\ea
 where $V_D(\pi_{c})$ represents the potential with an exact degeneracy.

Rewriting the potential (\ref{V2}) as a function of the thin-wall parameter $\epsilon$ yields \cite{Kolb:1988aj}
\ba
V(\pi_c)=\frac{\lambda}{4}\pi_c^2(\pi_c-\pi_c^D)^2-\frac{\lambda}{2} \e \pi_c^D \pi_c^3
 \label{V3}
\ea
This way of rewriting the potential uniquely define the thin-wall parameter $\e$ and introduces a new parameter $\pi_c^D$.
Note that this potential vanishes at the critical temperature so for $\pi_c=\pi_c(T_c)=\pi_c^D(T_c)$  since as we will show later, $\epsilon$ vanishes at the critical temperature .

Let us now turn to the definition of  $\pi_c^D$. In equation (\ref{pD}), $\pi_c^D$ was one minimum of the degenerated potential (so $V(\pi_c^D)=0$), the other minimum being at $\pi_c=0$.\\
Here the rewriting of the potential implies a new definition for $\pi_c^D$
  \ba
  \label{piD}
\pi_c^D (T)=2 \sqrt{\frac{D}{\la} (T^2-T_0^2)}= \frac{e}{\sqrt{3\la}}\sqrt{T^2-T_0^2}
\ea
It is interesting to check that when the potential is exactly degenerated, meaning when the temperature is the critical temperature, $\pi_c^D(T_c)$ matches the value found in (\ref{pD}).

\begin{figure}[h!]
\centering
\includegraphics[scale=0.7]{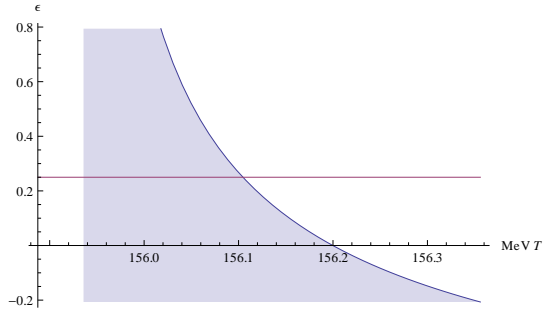}
\caption{Plot of the thin parameter $\e$, as a function of temperature for $\la=10 e^4$, $e^2=\frac{4\pi}{137}$, $\eta=66.47MeV$. These parameters correspond to $T_0=155.936$ MeV, and the critical temperature equals  $T_c=156.2$MeV.
The thin-wall approximation is valid only in a narrow energy range (less than $1$MeV) very close to $T_c$.
The horizontal line represents a $25\%$ threshold value (found in (\ref{pr})) above which the thin-wall approximation is not valid anymore.
For the thin-wall approximation to be valid, $\e$ must be much smaller than this threshold value.
As expected, the thin-wall parameter vanishes at the critical temperature and becomes negative above $T_c$.}
\label{tw}
\end{figure}

Likewise, the thin-wall parameter $\epsilon$ has now a new temperature dependent definition
\ba
\label{e1}
 \e(T)=\hspace{-1mm}\frac{ET}{\sqrt{\la D}} \frac{1}{\sqrt{T^2-T_0^2}}\hspace{-1mm}-\hspace{-1mm}1\hspace{-1mm}=\hspace{-1mm} \sqrt{\frac{T_c^2-T_0^2}{T^2-T_0^2}} \frac{T}{T_c}\hspace{-1mm}-\hspace{-1mm}1
=\hspace{-1mm}\sqrt{\frac{e^4}{3 \pi^2 \lambda} \frac{1}{1-\frac{T_0^2}{T^2}}}-1
 \ea
 This thin-wall parameter keeps track of the validity of the thin-wall approximation:
 the thin-wall approximation is only valid only when  $\e$ approaches zero. It is easy to see that our thin-wall parameter becomes negligible when the temperature approaches
 the critical temperature, $T_c$ . As shown in fig.\ref{tw} above $T_c$ the second minimum of the potential is higher than the initial one so that the field
is already in the true vacuum and no tunneling occurs: the thin-wall approximation does not make sense anymore and $\e$ becomes negative.
Far below $T_c$, the value of the thin-wall parameter explodes and the approximation breaks down.
As a consequence, the thin-wall approximation in only valid in the vicinity of $T_c$.

We just introduced a new potential form (\ref{V3}) with  2 new parameters : $\e$ and $\pi_c^D$.
In the following we study this new potential in one dimension and give preliminary results that will be useful to study bubble nucleation in 4D-euclidean spacetime.

First we can find the one dimensional action for the potential (\ref{V3})
\be
\label{S1}
S_1= \int dx [\frac{1}{2}(\frac{\partial_{\phi}}{\partial_x})^2+ V(\phi)]=\int_{\pi_c^D}^0 d\phi [2 V_D(\phi)]^{\frac{1}{2}}
=\frac{(\pi_c^D)^3 \sqrt{\la}}{6 \sqrt{2}}
\ee
 This action, like all the parameters in our theory, is temperature dependent
   \be
  S_1(T)=\frac{4}{3\sqrt{2} \la} [D (T^2-T_0^2)]^{3/2}=\frac{e^3}{18\sqrt{6} \la} (T^2-T_0^2)^{3/2}
\ee
Note that this action has  mass dimension 3 since it is the one-dimensional action. This one dimensional action is a necessary step to compute
the more complicated 4D euclidean action $S_{sphere}$. Indeed, $S_1$ directly appears in the final answer for $S_{sphere}$ in (\ref{S4}).
Physically, $S_1$ also corresponds to the bubble wall surface energy.
Another contribution to this 4D action comes from the potential energy density difference between the two minima
\ba
\label{DV}
\Delta V=V(\pi_{cmin})= \frac{\la}{2} \e (\pi_c^D)^4=8 (E T-\sqrt{\la D} \sqrt{T^2-T_0^2}) [\frac{D}{\la} (T^2-T_0^2)]^{\frac{3}{2}}
\ea
Here, contrary to the Mexican hat potential case (with no cubic terms) \cite{Coleman:1977py}, $\Delta V$, is not equal to the thin-wall parameter $\e$
which is a dimensionless quantity that decreases with temperature (see Fig.\ref{tw}).

Finally, for the particular shape of potential (\ref{V3}), one can easily find the solution of the differential equation (\ref{eq1})
and evaluate the thickness of the bubble wall. The solution for $\pi_c$ close to $r\sim R$ (where $R$ is the radius of the bubble wall) is \cite{Kolb:1988aj}
\ba
\pi_c(r) = \frac{1}{2} \pi_c^D [1-tanh(\frac{r-R}{\delta})]
\ea where $\delta$, the wall thickness, is equal to
 \ba
  \delta = \sqrt{\frac{8}{\la}}\frac{1}{\pi_c^D}= \frac{2 \sqrt{6}}{e} \frac{1}{\sqrt{T^2-T_0^2}} \label{del}
  \ea

As expected, the higher the temperature, the thinner the wall is. At lower temperatures the bubble wall becomes thick and the thin-wall approximation breaks down.

\subsection{Spherical instanton solution}
\begin{figure}[h!]
\centering
\includegraphics[scale=0.4]{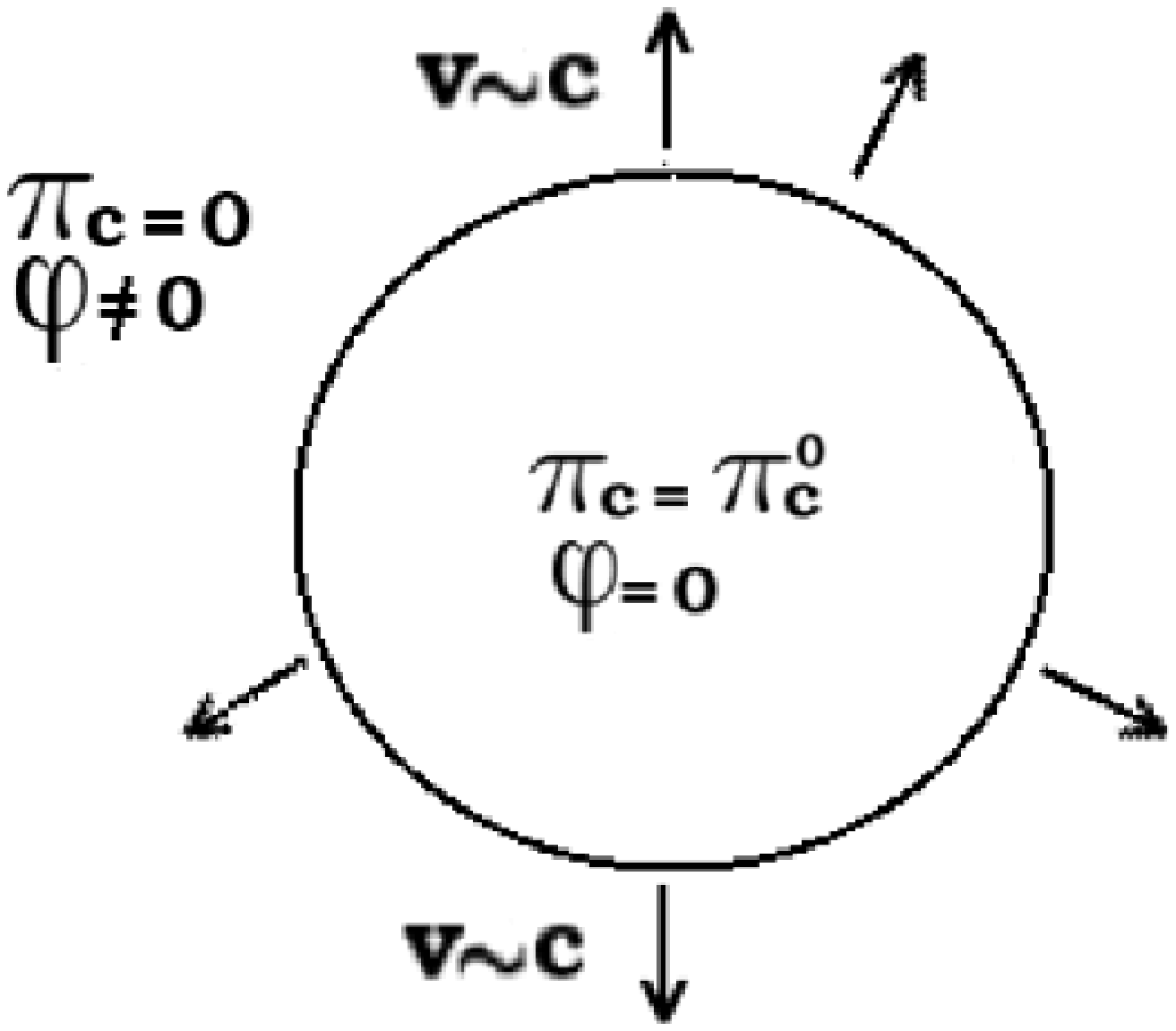}
\caption{Spherical shape for the instanton configuration\newline
In the idealized case, a sphere of radius R is generated and  expands radially at almost the speed of light $c$. Inside the bubble,
the charged field is in its true vacuum, $\pi_c^0$, and the neutral field vanishes (or is very close to zero as mentioned at the beginning of the discussion).
The arrows represent the way the bubble expands in Minkowski space.}
\label{br4}
\end{figure}

As mentioned earlier, the most probable way for the charged fields to tunnel through the potential barrier is through the nucleation of bubbles
with O(4) symmetry in Euclidean space (see \cite{Coleman:1977th} for proof).  The instanton is by definition a stationary solution in euclidean space and thus has a fixed 4-radius.
To get a physical picture one needs to Wick-rotate back to Minkowski space. When going to Minkowski space the configuration loses its stationarity and the physical radius of the bubble increases with time.
This bubble expansion is shown in fig.\ref{br4}. More precisely, after the bubble form, it will tend to expand at almost the speed of light (this will be justified later).

  However the above picture describes an idealized case where nothing prevents the bubble from growing.
  In our case, it is important to remember that the bubble forms in the core of a string which is itself embedded in a thermal plasma.
The bubble expansion will thus be slowed down or even stopped by  the plasma pressure and the string winding gradient. The expected effect on the string is that once the bubbles form, they get deformed and propagate along the string while melting its core.

Let us now take a closer look at the dynamics of the problem.
The equation of motion for a spherical bounce in 4-dimension
\ba
\frac{d^{2}\pi_{c}}{dr^{2}}+\frac{3}{r}\frac{d\pi_{c}}{dr}=V'(\pi_{c})
\ea
 which, as proved earlier, close to the critical temperature (so when the potential is almost degenerate) becomes
  \ba
\frac{d^{2}\pi_{c}}{dr^{2}}=V'(\pi_{c})\simeq V_D'(\pi_{c})
\ea

This equation is the same as the one yielding the 1 dimensional action (\ref{S1}).
Since the instanton is O(4) symmetric we now just have to integrate the lagrangian density over a 3-sphere of radius R to find the 4D euclidean action
\ba
\label{S4}
S_{sphere}&=\pi^2 \int r^3 d r [ \frac{1}{2} (\frac{\partial_{\pi_c}}{\partial_r})^2 + V(\pi_c)] \nonumber \\
           &=-\frac{\pi^2}{2} R^4 \Delta V +2\pi^2 R^3 S_1
\ea
where R is the radius of the bubble in Euclidean space.
The first term comes from the gradient energy associated with the bubble wall.
It is negative and grows like $\frac{\pi^2}{2} R^4$ which is the volume of the 4-sphere in euclidean space. The difference of potential energy  between the metastable and global minima $\Delta V$, also corresponds to the outward pressure.
The second term comes from the vacuum energy in the interior of the bubble:The area of the 3-sphere, $2\pi^2 R^3$ multiplies $S_1$, the surface tension of the bubble.
Since the volume of the bubble grows faster than its surface and the volume term in negative, the bubble loses potential energy when it expands.

In order to estimate the value of this euclidean action, we can extremize the above action and determine the radius $R(T)$ of the bubble nucleated.\\
The resulting radius reads
 \be R(T)= \frac{3 S_1}{\Delta V}=\frac{\sqrt{3}}{\sqrt{2} e} \frac{1}{\e\sqrt{T^2-T_0^2}} \label{r}\ee
 Making the temperature dependence of $\e$ manifest using (\ref{e1}), it is easy to see that $\frac{\partial_R}{\partial_T}$ is positive for any temperature below $T_c$. This means that the radius of the bubble increases with temperature.
As a result, when the Universe expands and cools down, the quantum tunneling  generates bubbles of smaller size. It is also interesting to evaluate the ratio of the wall thickness found in (\ref{del}) to the bubble radius
 \ba
 \frac{\delta}{R} = 4\e
 \ea
 For the thin-wall approximation $\e \ll 1$ so this ratio is, as expected, very small.

The final step in the computation of the dimensional euclidean action is to plug back the radius $R$ in (\ref{S4}).We can now read off the final answer for the 4D-euclidean action of the spherical instanton
\be
\label{S4b}
S_{sphere}=27 \pi^2 \frac{S_1^4}{2 {\Delta V}^3}=\frac{\pi^2}{48 \la \e^3}
\ee

Now, to obtain the bubble nucleation rate per unit volume  we simply exponentiate the above euclidean action
\ba
\frac{\Gamma_{sphere}}{V} \sim P_4 \exp[-\pi^2 \frac{1} {{48\ \la (\sqrt{\frac{e^4}{3 \pi^2 \lambda} \frac{T^2}{T^2-T_0^2}}-1)^3}}]
\label{ga}
\ea
The overall prefactor $P_4$ can be estimated on dimensional grounds as we will show in (\ref{PN}).
From the above equation, it is easy to see that the decay rate decreases with temperature and so the bubble nucleation rate increases when the temperature decreases.
As a result, the string should be less and less stable at lower temperatures when the thin-wall approximation applies.
It is important to remember that for temperatures too far below the critical temperature $T_c$, the thin-wall approximation breaks down.
In this case, we expect thick-wall bubbles with smaller actions and thus greater decay rates to form.

To end this section, let us describe the bubble after nucleation by computing the speed of the expanding wall together with its energy and its pressure.
The equation describing the stationary bubble in 4D-euclidean space is
\ba x^2+\tau^2=R^2 \ea
where R is the fixed 4-radius. Wick rotating back to Minkowski gives the equation of an hyperboloid
\ba x^2-t^2=R^2 \ea
So, in Minkowski spacetime, the physical radius of the bubble, $|\vec{x}|$, grows with time.
It is now easy to compute the velocity of the expanding wall after tunneling
\ba v=\frac{d|\vec{x}|}{dt}=\frac{\sqrt{|\vec{x}|^2-R^2}}{|\vec{x}|}\ea
  At some point, when $|\vec{x}|$ becomes much larger than R the velocity approaches the speed of light.
 However, in our case, the pressure the plasma exerts on the wall and the string winding gradient slow down this expansion.

Finally, the energy of the bubble wall in the string core depends on the wall velocity and the bubble surface energy $S_1$
 \ba E_{wall}= 4 \pi |\vec{x}|^2 S_1 (1-v^2)^{-\frac{1}{2}} \ea
Making the temperature explicit the energy of the wall becomes
\ba
\label{Ew}
E_{wall}=4 \pi |\vec{x}|^3 \frac{S_1}{R(T)}=
                  \frac {2  \pi \e}{27 \la}|\vec{x}|^3 e^4 (T^2-T_0^2)^2
                  \ea
As expected, when the temperature decreases, the energy of the wall decreases while the initial velocity of the bubble wall increases due to a smaller initial euclidean radius R.

Besides, the pressure of the wall bubble can naively be estimated from (\ref {Ew}) :
\ba P_{wall}=\frac{\vec{\nabla}E_{wall}}{4 \pi |x|^2}= \frac{3 S_1}{R}=\Delta V \ea
Note that this pressure does match the definition of the outward pressure appearing in the spherical euclidean action (\ref{S4}).
In order to estimate this wall pressure we simply need to estimate $\Delta V$.
From (\ref{DV}) we obtain
\ba P_{wall}= \Delta V= \frac{\la}{2} \e (\pi_c^D)^4= \frac{\e e^4}{18 \la^2} (T^2-T_0^2) \sim 0.005 \e (T^2-T_0^2)^2
\ea
We can now compare this pressure with the pressure exerted by the photon plasma: $$ P_{\gamma} \sim \frac{\pi^2}{135}T^4 $$
When the bubble nucleated is larger than the string radius, this plasma pressure slows down the wall expansion.
Comparing the pressure of the wall with the pressure of the plasma, we see that the photon plasma pressure dominates the wall pressure so we do not expect any expansion of the nucleated bubbles. However, inside the core of the string the plasma pressure is ineffective.
So we expect the plasma pressure to have an impact close to the critical temperature, when the bubbles nucleated are larger than the string's core.
When the temperature decreases, the pressure exerted by the thermal bath is lower and thus has a milder effect on slowing down radial expansion.

\subsection{Prefactor estimate}

 In the thin-wall regime, the tunneling rate is
determined mainly by the exponential part containing the euclidean action. However there is an overall prefactor $P_n$ that, in general, can be simply estimated using  dimensional analysis.
This prefactor is of order  \cite{Linde:1981zj}
 \ba \hspace{-1mm}P_n\hspace{-1mm}=\hspace{-1mm}\sqrt{(\frac {S_E^n }{2 \pi \hbar})}^n \hspace{-2mm}\sqrt{\frac{det[-\partial^2+ V''(\pi_c^*)] }
{det'[-\partial^2+ V''(0)]}}
\hspace{-1mm}\sim \hspace{-1mm}\sqrt{\frac {S_E^n }{2 \pi \hbar}}^n \hspace{-2mm} V''(\pi_c^*)^2
\label{PN}
\ea
where $det'$ denotes the determinant computed with the zero eigenvalue omitted
and $n$ corresponds to the number of zero modes associated with translational invariance in euclidean space \cite{Callan:1977pt}.
For the spherical case, $n=4$, for the cylindrical $O(3)$ case, $n=3$ and $n=2$ for the cylindrical $O(2)$ case mentioned in appendix \ref{ap}.
$S_E^n$ is the corresponding euclidean action for the $O(n$) symmetric configuration.
The prefactor computation for different symmetries of the instanton is more precisely discussed in \cite{Garriga:1994ut}\cite{Dunne:2005rt}.

To compute the prefactor in (\ref{PN}) we need both an estimate for the euclidean action, $S_E$ and for the second derivative of the potential.
Since we are in the thin-wall regime the temperature must be close to the critical temperature $T_c$ and the field values approaches $\pi_c^D$ defined in (\ref{piD}).

 We are now ready to estimate the prefactor (\ref{PN}).\\
 First we estimate the second derivative of the potential at $\pi_c \sim \pi_c^D$
\ba
\label{Ve}
 V''_{eff}(\pi_c^*) \sim V''_{eff}(\pi_c^D) \sim \frac{e^2 (T^2-T_0^2)}{6}
\ea
As one can see above, the second derivative of the potential is of order $T^2$ so that the total decay rate per unit volume has unit of $T^{4}$ as expected.\\
Second, we plug this result together with the spherical euclidean action (\ref{S4b}) in the expression for the spherical instanton prefactor  (\ref{PN})
 \ba
 P_4\hspace{-1mm} \sim \hspace{-1mm}\frac{3 \times 10^{-5} e^4}{\la^2 \e^6} (T^2-T_0^2)^2
  \ea
As one can see, this prefactor has a non trivial temperature dependence encoded in the thin-wall parameter $\e$.

A last step to know if this string can influence cosmology is to compare the value of the decay rate to the Hubble parameter.
Alternatively, we can also compare the decay rate per unit Hubble volume to the Hubble parameter.
The metastable string can have an impact on cosmology in any case. If it is quantum mechanically stable, it does not decay for a certain time until
the temperature drops below a certain temperature at which the string becomes topologically unstable.
If quantum tunneling occurs we expect the nucleation of bubbles which may also have an impact on cosmology.

 \subsection{Numerical results}

In order to give an order of magnitude for the decay we consider an example and study  our system at temperatures close to the QCD phase transition temperature.
The decay rate will in general be dominated by its exponential part (provided we use the thin-wall approximation).
As an example, let us use some typical values of the parameters  satisfying the constraints for $\la$
\ba
\la=10 e^4 \ \, \ e^2=\frac{4\pi}{137} \ , \ \eta=66.47 MeV
\ea
Here we chose $\la$ such that it satisfies the conditions found earlier in (\ref{lam1}), $\eta$ is the value in the true vacuum at zero temperature and $e$ is simply the electric charge.
For this choice of parameters, the critical temperature is $T_c=156.2$ MeV and the minimum temperature is $T_0=155.936$ MeV.
Using the action given by (\ref{S4}) we can now evaluate the bubble nucleation rate for different values of the thin-wall parameters : \\
For example, for a thin-wall parameter equal to a tenth of a percent($\e \sim 0.1\%$), the decay rate per unit volume is
  \ba
  \label{d1}
  \frac{\Gamma}{V} \sim P_4  e^{-2.4 .10^9}
  \ea
Whereas when the thin-wall parameter is of order of a few percents ($\e \sim 3\%$), the decay rate is much bigger
\ba
\label{d2}
\frac{\Gamma}{V} \sim P_4 e^{-9.0 .10^4}
\ea
However, we should not forget that $\e$ must be small for the thin-wall approximation to remain valid so we expect the first example to be more accurate than the second one.
Now that we found the decay rate in Minkowski space time, we can compare it to the Hubble parameter H in order to know if quantum tunneling actually occurs in an expanding universe. The string will be stable under quantum breaking only when the decay rate per Hubble volume
is much smaller than the Hubble parameter
\be
\frac{\Gamma}{V} \ll H^4 \
\ee
where $\ H \sim \frac{1}{t} $.

As an example, for the QCD phase transition chosen above, the typical time is $t_{QCD}\sim 10^{-5}s $  and the corresponding Hubble volume is  $H^4 \sim 10^{-20}s ^{-4}\sim e^{-46}s ^{-4} $. By comparing the Hubble volume to the decay rate found in (\ref{d1}) and (\ref{d2}),
we see that in both cases the decay rate per unit volume is much smaller than $H^4$.
As a result, we expect this string to be truly stable against tunneling in the charged field direction during the  QCD phase transition.
Below the critical temperature, the thin-wall approximation does not hold anymore and we expect thicker bubbles with smaller action.
We therefore expect a larger decay rate when the temperature decreases and the string becomes quantum mechanically unstable
below a certain temperature.

\section{Discussion}
In the following we comment on the various approximations we have made to describe the neutral string in contact with a thermal bath of photons.

First, in order to estimate the decay rate of the string, we have used the semiclassical thin-wall approximation which only applies when the instanton action is very large, $S_E \gg 1$.
In addition, this approximation holds only if  the radius of the bubble wall is much bigger than its thickness
\ba
\label{pr}
\frac{3 S_1}{\Delta V} \gg \sqrt{\frac{8}{\la}}\frac{1}{\pi_c^D}
\ea
or equivalently, $ \e \ll \frac{1}{4} \ \ or \ \ T \gg \frac{1}{1-\frac{4 e^4}{15 \la \pi^2}} \label{small}$.
As expected, condition (\ref{pr}) imposes a very small thin-wall parameter.

Second, coming back to our initial string picture it is important to remember that we considered tunneling only in the core of the string since it is where the energy is localized. This assumptions means that the neutral fields almost vanishes throughout quantum tunneling.
But in reality, the tunneling may happen in the $\phi$ field direction as well.
 To check this assumption, let us consider both the charged field $\pi_c$ and the neutral field $\phi$ after tunneling.
 For the full configuration to be in the true vacuum it is necessary to minimize the full potential $V(\phi,\pi_c)$.
So, if the $\pi_c$ field takes on its vacuum expectation value $\pi_c^-(T)$, we expect $\phi$ to become $\phi-(T)= \sqrt{\eta^2-\pi_c-}$
to minimize the quartic term of the potential. As a result, $ \phi$ becomes a non zero function of temperature inside the bubble.
As a typical value for the charged field after tunneling we choose $\pi_c-=\pi_c*(T_c)$ so that the thin-wall approximation is valid.
This value of $\pi_c-$ yields a specific value for the neutral field $\phi = \sqrt{\eta^2-\pi-^2}$ which is much smaller than one.
Therefore we proved that the approximation $\phi \sim 0$  during the quantum tunneling is a consistent approximation although
it may not capture all the subtleties of the tunneling.

Let us now compare the energy per unit length within the core of the string before and after tunneling.
Before tunneling $\pi_c=0$, so that the string configuration looks simply like the global $U(1)$ neutral string at zero temperature.
There is no gauge field to compensate the gradient energy so that the total energy is divergent.
 \ba
E \sim 2\pi \eta \ln(\frac{L}{w}) + \frac{\pi \eta^2}{2}
\label{Esa}
\ea
where $w = \frac{1}{\sqrt{\la}\eta}$ is the width of the string and L the cutoff scale.
The first term comes from the winding around the neutral string and the second corresponds to the energy trapped in the core of the string.
However, for the neutral string this energy divergence is mild (only logarithmic) and a cutoff can be implemented in cosmological context  (for example the nearest string).
After tunneling, the energy per unit length within the core of the static string depends on temperature
\ba
E \hspace{-1mm}\sim \hspace{-1mm} \pi w ^2 [{\frac{\lambda}{4}((\pi_c-)^2-\eta^2)^2+\frac{e^2}{12}T^2 (\pi_c-)^2-\frac{e^3}{6\pi}T(\pi_c-)^3}]  \nonumber
\label{Esb}
\ea
where $\pi_c-$ is the value of the charged field in the true vacuum.
As expected, the energy of the core is lower after tunneling. More importantly, once formed, the string will classically dissolve at zero temperature since, at zero temperature the vacuum manifold is a 3-sphere.

Another important assumption we made in our picture is about the temperature dependance of the fields. In our case, the scalar fields are non thermal although there are coupled to a thermal bath.
The resulting effective potential tells us that inside the core of the neutral string, the energy is maximal but the state can quantum mechanically decay to a lower energy state. The result is that the core of the string can still decay to the true vacuum by tunneling in the charged field direction. This decay in the core of the string occurs by the nucleation of the most probable instanton configuration :  bubbles with O(4) symmetry in euclidean space are nucleated.

In contrast, if all the fields were in thermal equilibrium the decay would happen via a purely thermal process:
The charged field can hop over the barrier and nucleate a static instanton with O(3) symmetry instead of the O(4) we found.
This thermal process nucleates instantons called sphalerons which can dominate the whole quantum tunneling process only if the
scalar fields are in thermal equilibrium.
In our case, since we consider out-of-equilibrium scalar fields, such a sphaleron production does not occur.

Let us finally discuss what happens during the three different temperature ranges, below the critical temperature $T_c$, closer to $T_0$ and below $T_0$ : 

Close but below the critical temperature, where the thin-wall approximation is valid, the value of the true vacuum, $\pi_c-$, decreases when temperature is lowered.
 As a result, the bubble wall thickness, $\delta \sim \sqrt{\frac{8}{\la}}\frac{1}{\pi_c-}$ increases.
 In addition, since $\pi_c-< \eta$ the thickness of the string is always greater than the core of the string.
In the thin-wall regime, we also expect the radius to be much bigger than the bubble wall thickness.
 Consequently, the radius of the nucleated bubble is larger than the core of the string.
As a result, the string breaks at points where tunneling occurs. For example, this is what happens when a string breaks by nucleating a pair of monopoles.

At lower temperatures, closer to $T_ 0$, the height of the barrier of the potential becomes smaller than its depth:
 One of the potential minima is much lower than the other and the wall becomes thicker.
As a result, the total potential energy contained inside the bubble is smaller and the surface term dominates.
Moreover, we expect a smaller action. So the absolute value for the argument in the exponential part of the decay rate is small and the exponential goes to 1. As a result, the prefactor includes most of the relevant information for the decay rate and we still expect a bigger decay rate for thick-wall than for thin-wall bubbles.
Consequently, at small enough temperature, the larger bubble nucleation rate make the string unstable.
Besides, in this case,  the tunneling happens within the core of the string. The bubble has a tendency to extend in all directions
but the neutral string gradient prevents any radial expansion. The instanton grows until it reaches approximately the radius of the string, then
gets deformed due to the string gradient. Finally it propagates along the string, with the configuration inside the bubble being close to the true
vacuum. In other words, it melts the core of the string where all the energy was initially stored.

Finally, below $T_0$, the charged field decays classically so that the string is initially unstable. We are back to the zero temperature
case: the vacuum manifold is $S_3$ and the string is topologically unstable.

\section{Conclusion}

We have found new results for the decay rate of the neutral string in contact with a thermal bath of photons.
We have studied quantum tunneling for  specific string configurations with two complex fields : one charged coupled to a photon bath
and one neutral making up the string.
We applied the thin-wall approximation to compute the radius of the instanton after quantum tunneling, the energy of the bubble wall and the speed of bubble expansion with their new temperature dependence.

For the Landau-Ginzburg string we study, the potential is metastable and  the following results hold for strings in contact with a thermal bath  :
\begin{itemize}
\item The string can be stabilized by a thermal bath of photons even if all the scalar fields considered are out of equilibrium.
\item Two main effects prevent the string from decaying into radially expanding configurations: plasma pressure and the string winding gradient, which creates the energy barrier enclosing the core of the string.

  \item Below but close to $T_c$, the Landau- Ginzburg string is effectively stable against tunneling since the decay rate per unit volume (\ref{ga}) is much smaller than the Hubble parameter.
Indeed, for a certain range of temperature below $T_c$,  $\frac{\Gamma}{V} \ll H^4$.
\item  The radius of the bubble nucleated is (\ref{r}) and increases with temperature .
\item  Inversely, the decay rate of the string (\ref{ga}) decreases with temperature, so the string is more stable at high temperatures.
 \item At temperatures close to the critical temperature $ T_c$, the bubble radius is bigger than the neutral string typical thickness, so that it effectively breaks the string.

 \item At lower temperature  bubbles nucleated have a smaller size, so that the decay can happen within the core of the string .

After nucleation, bubbles expand until they reach the radius of the string and then get deformed. Meanwhile,  the expansion of each bubble makes the string thicker as it tries to expand at almost the speed of light. Thus, after nucleation, instantons propagate along the string and the energy trapped in the core of the string goes to a true vacuum.
In that sense, instanton production makes the string melt and breaks the string at the point where it occurs.

\end{itemize}

Below the temperature $T_0$, the charged field classically rolls down the potential and the string
  becomes topologically unstable. As a result, the string is metastable only for a certain range of temperature between $T_0$ and $T_c$.

   Interestingly, the nucleation of instantons in our setup implies the existence of non-trivial field configurations called skyrmions.
    Opposite winding direction configurations for the charged field components form skyrmion anti-skryrmion pairs when they meet. If the two configurations have the same alignment, they pass through each other and no skyrmions form. When skyrmions form, they can spread outside the strings since cosmic strings are in fact moving.

    Finally, in a thermal bath of photons, the new melting effect due to the tunneling in the charged field direction can destabilize the classically stable string. After nucleation and below $T_c$, we expect the thick-wall bubbles to be deformed and to propagate along the string at almost the speed of light. As a result, the string core melts and the two newly formed string segments move away from each other at the almost the speed of light. The melting is more likely to occur far below the critical temperature, when the string decay rate is large enough compared to the Hubble parameter.

\begin{acknowledgments}I wish to thank Robert Brandenberger and Ajit Mohan Srivastava for  discussions during this project and comments on the draft .
I would also like to thanks to Mark Hertzberg and Mark Kon for their help with proofreading.
This work was supported in part by the U.S. Department of Energy under cooperative research agreement DE-FG02-05ER-41360. The author is supported by an NSERC PDF fellowship.\end{acknowledgments}
\appendix
\section{String Instanton nucleation at zero temperature}
\label{ap}
In the following we describe a possible ansatz for the string instanton at zero temperature and give only qualitative results.

Instantons are classical solutions to the euclidean equation of motion with non-zero action.
In order to study the quantum stability of the string against  breaking, one can find the instanton configuration
which corresponds to the tunneling from the false vacuum to the true one \cite{Coleman:1977py}.
The resulting decay rate is then
\ba
\Gamma \propto e^{-(S_E-S_0)}
\ea
where $S_E $ is the euclidean action of the instanton and $S_0$ is the background string action.
To describe the instanton mediating the breaking of the string, we first introduce an ansatz for the cylindrical instanton. This cylindrical instanton seems to be the most natural
one since it has the same rotational symmetry around the string axis as the string itself. It is therefore necessary for us to work with the cylindrical coordinates ($\rho$,$\theta$ and z where $\rho$ is the radial distance
to the string axis, $\theta$ the azimuthal angle and z  the height).

However, it is important to remember that other configurations with a different symmetry may be nucleated as well: For example, a string can decay through the nucleation of a monopole-anti monopole pair \cite{Preskill:1992ck}.
Those monopoles are spherically symmetric and so also have the required rotational symmetry around the string. In the case where monopole-anti-monopole pairs are nucleated, we expect the speed of the bubble expansion to be further reduced.

\begin{figure}[htbp]
\centering
\vspace{-0.2cm}
\includegraphics[scale=0.4]{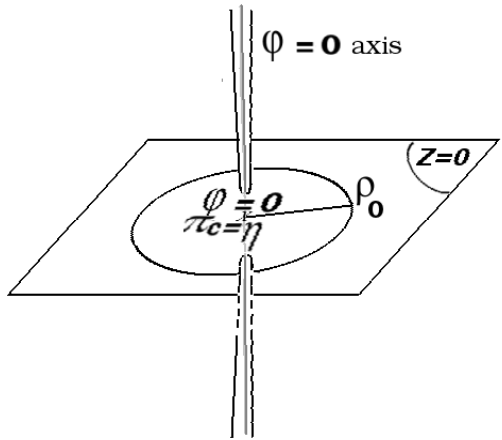}
\vspace{-0.2cm}
\caption{Breaking of a string into two strings at zero temperature.
The vertical axis represents the core of the string lying along the z-axis.
In the $z=0$-plane, the charged field takes on its v.e.v. $\langle \pi_c \rangle=\eta$,
inside the disk of radius $\rho_0$ centered on the string. In this region the field $\phi$ vanishes.}
\label{br2}
\end{figure}

Let us now describe the string-breaking mechanism. The initial neutral string is simply a global U(1) string and lies on the z-axis.
After breaking, the string nucleates an instanton within which the complex charged field takes on its vacuum expectation value. This string breaking through instanton
nucleation is shown in fig.\ref{br2}. Here the picture describes the nucleation of an instanton greater than the core of the string. This large instanton would instantaneously break the string.
We choose the string-breaking instanton to be centered around the $z=0$-plane.

In our set-up the initial configuration corresponds to infinite times $t=\pm \infty$ and the tunneling occurs at $t=0$
  \ba
  t: &\pm\infty  \rightarrow 0 \nonumber \\
      &(\phi_i, \pi_{ci})   \rightarrow   ( \phi_b , \pi_{cb})
\ea

At $t=\pm \infty$ the straight static string configuration with cylindrical symmetry can be written as
\ba
\label{c1}
(\phi_i , \pi_{ci})=( \eta  f (\rho) e^{in\theta} ,0)
 \ea
 where $\rho$ is the distance to the core of the string, $n$ is the winding number and $\theta$ is the angle in the plane orthogonal to the string.
 $f(\rho)$ is the standard real function representing a $U(1)$ global string which interpolates between 0 and 1:  $f(0)=0$ and $f(\infty)=1$ .

\begin{figure}[h!]
\centering
\includegraphics[scale=0.4]{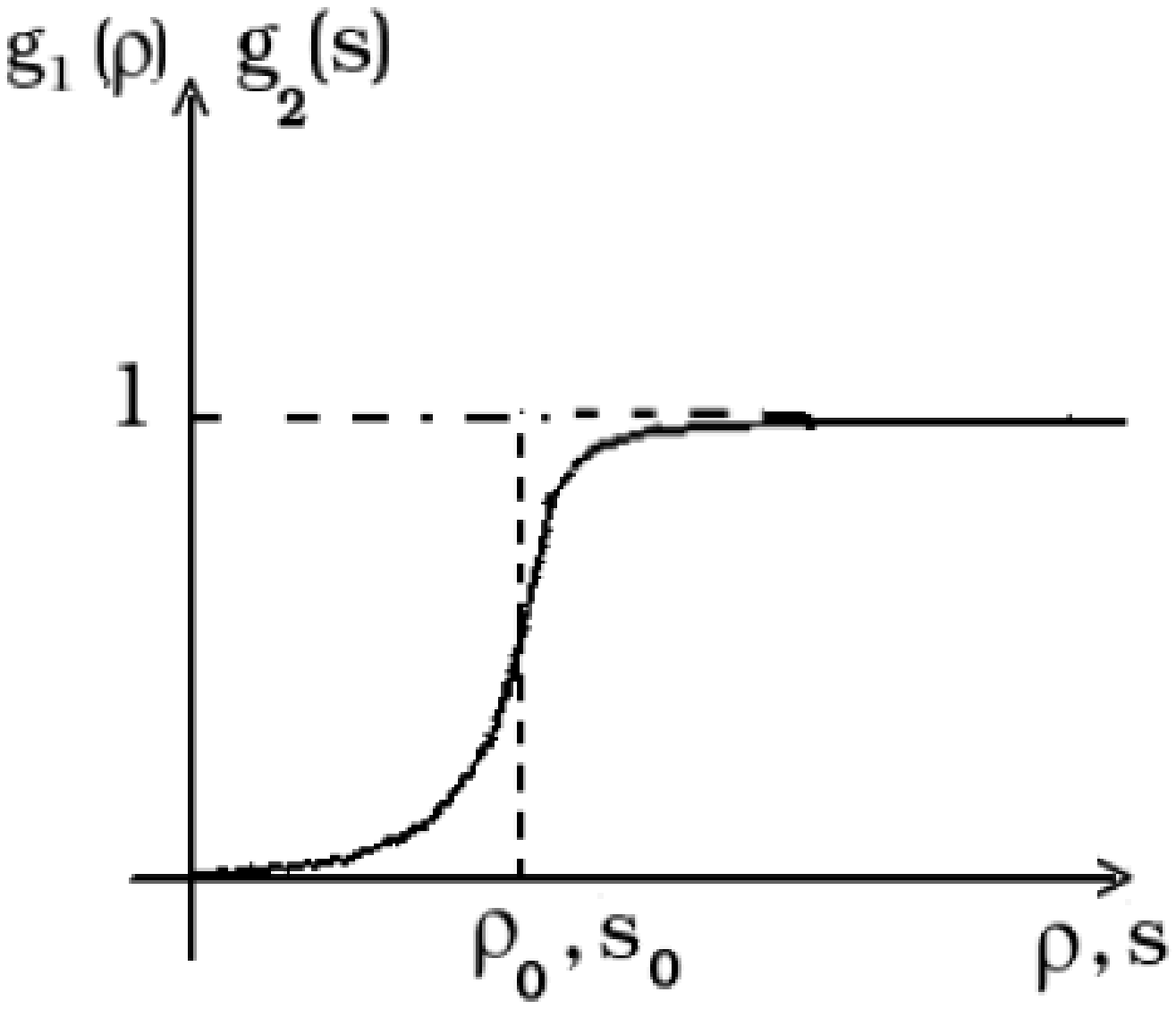}
\caption{Sketch of the profile function for $g_1(\rho)$ and for $g_2(s)$\newline
The profile function for $g_1(\rho)$ and $g_2(s)$ look similar and interpolate smoothly between the initial and the final string configuration.}
\label{br3}
\end{figure}

After tunneling, the string breaks and the two half-strings extend on each side
of the z=0 plane.This breaking happens at $t=0$ where the instanton solution ‚Äúbounces‚Äù.
Afterwards the solution asymptotically reaches the initial state $(\phi_i,\pi_{ci})$ at $t=+ \infty$ \cite{Coleman:1977py}.
As shown in Fig.\ref{br2}, after tunneling
the field $\phi$ vanishes on the $z=0 $-plane by continuity in the vicinity of the z-axis since it also vanishes in the core
of the initial and final strings.
In order to keep the cylindrical symmetry of the system, we choose the patch where $\phi=0$ to be a disk
of radius $\rho_0$ centered on the origin. Roughly speaking the resulting configuration can be described by the following ansatz
 \ba
\label{c2}
\phi_b = \begin{cases}
\eta   f (\rho) e^{in\theta}  \ \ \ for  \ z \neq 0 \  and \  {z = 0 , \rho \geq \rho_0}  \\
\sim 0                  \ \ \             for \ z=0  \ and \ \rho < \rho_0
\end{cases}
\ea
The real solution smoothly interpolates between the $\phi(\rho < \rho_0)$ and  $\phi (\rho > \rho_0)$ as shown in fig.\ref{br3}.
In the thin-wall approximation, the interpolation is very steep.

Considering tunneling only in the core of the string, one can assume that $\phi \sim 0$ at any time.
$\rho_0$ is simply the radius of the instanton so that outside of the instanton, at radius larger than $\rho_0$, the string
configuration is the usual one: $\phi(\rho,\theta)=\eta   f (\rho) e^{in\theta}$.

For the whole configuration to lie on the vacuum manifold, the charged fields have to be turned on for the patch defined above
 \ba
 \label{c3}
 \pi_{cb}= \begin{cases}
 \eta \ for \ {z=0, \ \rho < \rho_0}  \\
\sim 0 \ for \ z=0, \ \rho > \rho_0  \ \ and\ \  for \ z\neq 0 \
\end{cases}
\ea

Again this is simply a schematic ansatz, and the real solution should smoothly interpolate between  $\pi_c(\rho < \rho_0)$ and  $\pi_c (\rho > \rho_0)$ (see fig.\ref{br3} for a sketch of the profile functions describing the instanton solution).

In order to find the instanton we have to work in euclidean space. To do so we wick rotate the time and introduce a new Euclidean time variable $\tau=i t$.
It is important to note that despite the fact we have a temperature-dependent potential, the Euclidean time is not periodic with
period $\frac{\hbar}{k_b T}$  since all the scalar fields we are dealing with are non-thermal.
The temperature $T$ only comes from the thermal bath of photons to which the non-thermal fields are coupled.

\begin{figure}[h]
\centering
\includegraphics[scale=0.5]{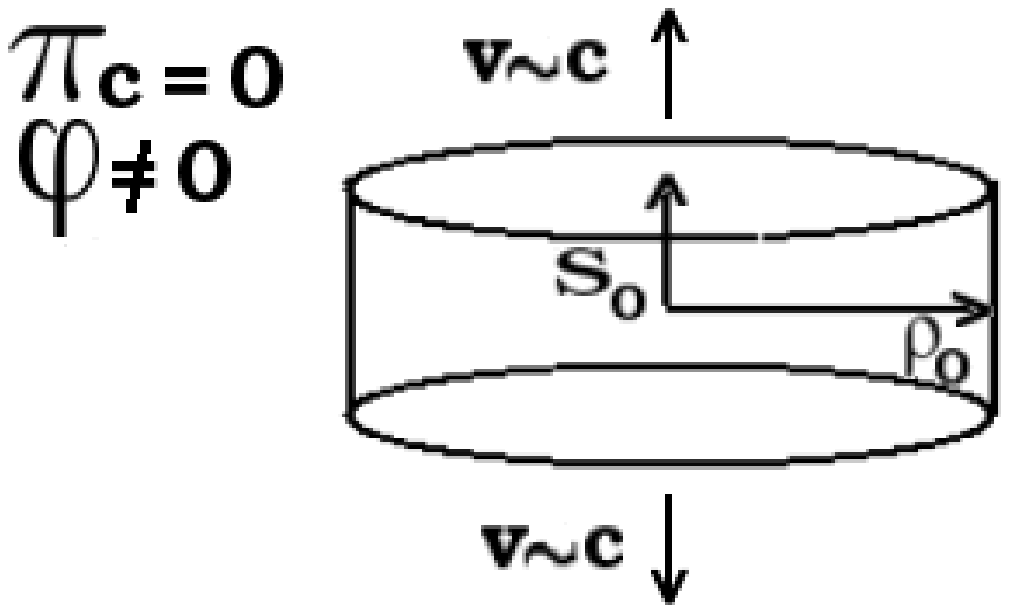}
\caption{Idealized cylindrical shape for the instanton configuration\newline
When one subtracts the bounce configuration from the original string, one is left with
a cylinder of radius $\rho_0$ and height $s_0$ that is expanding vertically, along the z-axis, at the speed of light c.
The arrows represent the way the cylinder expands in Minkowski space.}
\label{puck}
\end{figure}

In order to describe a cylindrical instanton expanding along the string axis we introduce a new coordinate that mixes
the z-axis and the euclidean time coordinates :   $$s=\sqrt{\tau^2+z^2}$$
 This instanton configuration now has two O(2) symmetries. One in the plane orthogonal to the string corresponds to a cylindrical symmetry
and is described by the ($\rho, \theta$) coordinates.
The other lies in the ($z-\tau$)-plane.  Rotating back to Minkowski spacetime yields one O(1,1) symmetry in the (z,t)-plane and one O(2) in the plane orthogonal to the string.
This means that, as shown in fig.\ref{puck} the nucleated cylinder expands along the string. After tunneling we expect the edges of the cylinder to be smoothed out.

We now have to describe the instanton solution which interpolates between $\phi_i$ and $\pi_{ci}$ and $\phi_b$ and $\pi_{b}$.
In order to do so, we introduce two new functions: the first one, $g_1(\rho)$ depends uniquely on $\rho$, and the other one, $g_2(s)$ uniquely depends on the new variable $s$.
In terms of those two functions, the expanding cylindrical instanton anzatz becomes
\ba
\label{a4}
\phi_(\tau,\rho,z,\theta)  && \hspace{-0.2cm}= \eta f (\rho) e^{in\theta} [g_1(\rho)+\sqrt{1-g_1(\rho)^2}g_2(s)] \\
\pi_c(\tau,\rho,z,\theta) && \hspace{-0.2cm}=\eta \sqrt{1-g_1(\rho)^2}\sqrt{1-g_2(s)^2}
\label{a5}
\ea

The field value interpolates from close to the true vacuum at $s\rightarrow 0$ to exactly the false vacuum as $s \rightarrow \infty$. In the thin-wall limit this profile sharpens to a step function.
Note that this solution has two O(2) symmetries:
one in  $(z,\tau)$-plane and one in the $(\rho,\theta)$-plane, so that $\phi_(\tau,\rho,z,\theta)=\phi_(\rho,s,\theta)$.

Let us now impose the general boundary conditions to obtain a bounce to find the shape of the two new functions $g_1(\rho)$ and $g_2(s)$.
One important condition is that the derivative vanish at the bouncing point (when $\tau=0$)
\ba
\label{bc}
\partial_\tau (\phi(\tau), \pi_c(\tau))_{\tau=0}= \partial_\tau (\phi_b, \pi_{cb})=(0,0)
\ea
The second condition simply corresponds to the fact that at infinite
times (when $\tau=\pm \infty$) the field takes on its initial configuration
\ba
\label{bc2}
\lim_{\tau \rightarrow \pm \infty }  (\phi,\pi_c) =(\phi_i,\pi_{ci})=(0,0)
\ea

Together with (\ref{a4})(\ref{a5}), the boundary conditions (\ref{bc})(\ref{bc2}) yield
\ba
g_1(0)=g_1(\rho \leq \rho_0)=0 \ , g_1(\rho > \rho_0)=1\ , g_1'(0)=0 \\
g_2(0)=0 \ , g_2(\pm \infty)=1 \ ,  g_2'(0)=0
\label{grad}
\ea
where ' denotes the derivative with respect to $s$ (respectively $\rho$), since  (\ref{bc}) holds at any point of space and we can thus replace
$\tau$ by $s$ (respectively $\rho$).

The above constraint allows us to sketch the profile for the two functions $g_1(\rho)$ and $g_2(s)$.
As shown in fig.\ref{br3}, the initial value of $g_1$ asymptotically vanishes at small $\rho$ and then jumps at $\rho \sim \rho_0$ .
Above the threshold value $\rho_0$, the function goes to 1. This amounts to having the charged fields turned on only inside
the disk of radius $\rho_0$ at the bouncing point $(t=0)$.
The other profile function $g_2$ has the same shape as $g_1$ although the jumping does not necessary occur at the same value ($s_0 \neq \rho_0 $).
The initial value of $g_2$ vanishes since $\phi(0)=0$ and has a vanishing derivative at the bouncing point (see (\ref{grad})).
It then jumps to 1 to come back to the single neutral string configuration at $t= + \infty$. The field value interpolates from close to the true vacuum at $s\rightarrow 0$ to exactly the false vacuum as $s \rightarrow \infty$. Thus it describes well our instanton representing the tunneling between the initial one string configuration and the final two half-string configuration.

It is interesting to note that we asymptotically reach  the value 1 for the profile of each of the three functions $f, g_1$ and $g_2$.
In the thin-wall limit the profile of the two functions $g_1(\rho)$ and $g_2(s)$ sharpens to a step function.

In order to find the actual value for $\rho_0$ and $s_0$, one would have to extremize the euclidean action $S_E$ to find where the saddle point is.
This imposes $\frac{\partial_{S_E}}{\partial_{\rho_0}}=0$ and $\frac{\partial_{S_E}}{\partial_{s_0}}=0$.

Let us now comment on the feature of quantum tunneling in the string-breaking case at zero temperature. Tunneling from the false vacuum to the true vacuum proceeds here by cylinder nucleation.
In the semiclassical description the field makes a quantum jump in Minkowski spacetime: a zero-energy bubble of true vacuum  appears in the false vacuum. The bubble classically grows and completes the transition.

Let us evaluate the typical length/time scale of the quantum tunneling in Minkowski spacetime.
In order to go from the initial configuration $(\phi_i,\pi_{ci})$ to the bouncing one, $(\phi_b,\pi_{cb})$, the fields jump quantum mechanically during a very short
transition time. One can estimate the typical length scale where this occurs using the Heisenberg uncertainty principle in Minkowski spacetime: $\Delta E \Delta t\sim \hbar  $.
 The typical energy we consider is the energy of the core of the string: $E\sim \frac{\la}{4} \eta^4 V$ where $V$ is the volume.
 Assuming the tunneling happens within the string core, the maximal volume of a cylinder of heigt $2 s_0$ is $V \sim 2 \pi w^2 s_0$ where $w \sim (\sqrt{\la}\eta)^{-1}$ is the width of the string.
 We therefore obtain the typical tunneling length, $l \sim t \sim \frac{2 \sqrt{\lambda }}{\pi \eta^2 s_0}$ in units where $\hbar =c=1$.
For $s_0$ big enough,  this typical length $l$ is smaller than the string radius $l < w$ and, assuming that tunneling occurs within the core of the string, is a valid approximation.
We also expect this $l$ to roughly correspond to $\rho_0$ since both describe the typical length at which the solution transits from the false vacuum to the true vacuum.
Here $s_0$ corresponds to the point where the quantum jump occurs, meaning that we can infer that the profile of $g_2(s)$ has a sharp
but continuous change at $s=s_0$. In a similar manner, we have already mentioned that the profile of $g_1(\rho)$ has a sharp change at $\rho=\rho_0$  (see Fig.\ref{br3}).

The effect of the quantum tunneling is to divide the string in two at the tunneling point. Indeed, the boundary of the cylindrical region separating the two strings expands along the string, keeping the radius $\rho_0$ fixed
        \ba z^2-t^2= s_0^2 \ea
As a result we see that the expansion of this cylindrical wall is purely along the z-direction.
  From the above equation one can deduce the expansion velocity of the cylinder
   \ba v=\frac{dz }{dt}= \frac{\sqrt{z^2-s_0^2}}{z} \ea
   Since $s_0$ is a constant and $z$ grows with time, this velocity rapidly becomes the speed of light.
As a consequence, the two newly formed strings quickly move away from each other along the z-axis at the speed of light, while the core of the string melts.


\begin{thebibliography}{99}

\bibitem{ShellVil}
A. Vilenkin and E.P.S. Shellard, \textit{Cosmic Strings and other
Topological Defects} (Cambridge Univ. Press, Cambridge, 1994).

\bibitem{HK}
M.~B.~Hindmarsh and T.~W.~B.~Kibble,
  ``Cosmic strings,''
  Rept.\ Prog.\ Phys.\  {\bf 58}, 477 (1995)
  [arXiv:hep-ph/9411342].

\bibitem{RHBrev}
R.~H.~Brandenberger,
  ``Topological defects and structure formation,''
  Int.\ J.\ Mod.\ Phys.\ A {\bf 9}, 2117 (1994)
  [arXiv:astro-ph/9310041].
  
\bibitem{Brandenberger:1994bx}
  R.~H.~Brandenberger, A.~-C.~Davis and M.~Trodden,
  Phys.\ Lett.\ B {\bf 335}, 123 (1994)
  [hep-ph/9403215].

\bibitem{CSmag}
R.~H.~Brandenberger and X.~-m.~Zhang,
  ``Anomalous global strings and primordial magnetic fields,''
  Phys.\ Rev.\ D {\bf 59}, 081301 (1999)
  [hep-ph/9808306].

\bibitem{Tashiro:2012pp}
  H.~Tashiro, E.~Sabancilar and T.~Vachaspati,
  arXiv:1212.3283 [astro-ph.CO].

\bibitem{Bevis:2007gh}
  N.~Bevis, M.~Hindmarsh, M.~Kunz and J.~Urrestilla,
  Phys.\ Rev.\ Lett.\  {\bf 100}, 021301 (2008)
  [astro-ph/0702223 [ASTRO-PH]].



\bibitem{Bhattacharjee:1998qc}
  P.~Bhattacharjee and G.~Sigl,
  Phys.\ Rept.\  {\bf 327}, 109 (2000)
  [astro-ph/9811011].

\bibitem{Nielsen:1973cs}
  H.~B.~Nielsen and P.~Olesen,
  Nucl.\ Phys.\  B {\bf 61}, 45 (1973).
\bibitem{Halperin:1973jh}
  B.~i.~Halperin, T.~C.~Lubensky and S.~-k.~Ma,
  Phys.\ Rev.\ Lett.\  {\bf 32}, 292 (1974).





\bibitem{Linde:1981zj}
  A.~D.~Linde,
  Nucl.\ Phys.\  B {\bf 216}, 421 (1983)
  [Erratum-ibid.\  B {\bf 223}, 544 (1983)].



\bibitem{Quiros:1999jp}
  M.~Quiros,
  arXiv:hep-ph/9901312.

\bibitem{Kibble:1976sj}
  T.~W.~B.~Kibble,
  J.\ Phys.\ A  {\bf 9}, 1387 (1976).


\bibitem{Dolan:1973qd}
  L.~Dolan and R.~Jackiw,
  Phys.\ Rev.\  D {\bf 9}, 3320 (1974).

\bibitem{Dine:1992wr}
  M.~Dine, R.~G.~Leigh, P.~Y.~Huet, A.~D.~Linde, D.~A.~Linde,
  Phys.\ Rev.\  {\bf D46}, 550-571 (1992).
  [hep-ph/9203203].\\
  M.~Dine, R.~G.~Leigh, P.~Huet, A.~D.~Linde, D.~A.~Linde,
  Phys.\ Lett.\  {\bf B283}, 319-325 (1992).
  [hep-ph/9203201].




\bibitem{Holman:1992rv}
  R.~Holman, S.~Hsu, T.~Vachaspati and R.~Watkins,
  Phys.\ Rev.\  D {\bf 46}, 5352 (1992)
  [arXiv:hep-ph/9208245].
  
\bibitem{Dasgupta:1997kn}
  I.~Dasgupta,
  Nucl.\ Phys.\  {\bf B506}, 421-435 (1997).
  [hep-th/9702041]

\bibitem{Garriga:1991tb}
  J.~Garriga and A.~Vilenkin,
  Phys.\ Rev.\  D {\bf 45}, 3469 (1992).


\bibitem{Preskill:1992ck}
  J.~Preskill and A.~Vilenkin,
  Phys.\ Rev.\  D {\bf 47}, 2324 (1993)
  [arXiv:hep-ph/9209210].


\bibitem{Monin:2008mp}
  A.~Monin and M.~B.~Voloshin,
  Phys.\ Rev.\  D {\bf 78}, 065048 (2008)
  [arXiv:0808.1693 [hep-th]].

\bibitem{Monin:2008uj}
  A.~Monin and M.~B.~Voloshin,
  Phys.\ Rev.\  D {\bf 78}, 125029 (2008)
  [arXiv:0809.5286 [hep-th]].


\bibitem{Karouby:2012yz}
  J.~Karouby and R.~Brandenberger,
  Phys.\ Rev.\ D {\bf 85}, 107702 (2012)
  [arXiv:1203.0073 [hep-th]].





\bibitem{Nagasawa:1999iv}
  M.~Nagasawa and R.~H.~Brandenberger,
  Phys.\ Lett.\ B {\bf 467}, 205 (1999)
  [hep-ph/9904261].


 


\bibitem{Coleman:1977py}
  S.~R.~Coleman,
  Phys.\ Rev.\  D {\bf 15}, 2929 (1977)
  [Erratum-ibid.\  D {\bf 16}, 1248 (1977)].

\bibitem{Arnold:1992rz}
  P.~B.~Arnold and O.~Espinosa,
  Phys.\ Rev.\ D {\bf 47}, 3546 (1993)
  [Erratum-ibid.\ D {\bf 50}, 6662 (1994)]
  [hep-ph/9212235].

\bibitem{Kolb:1988aj}
  E.~W.~.~Kolb and M.~S.~.~Turner,

\bibitem{Coleman:1977th}
  S.~R.~Coleman, V.~Glaser and A.~Martin,
  Commun.\ Math.\ Phys.\  {\bf 58}, 211 (1978).


\bibitem{Callan:1977pt}
  C.~G.~.~Callan and S.~R.~Coleman,
  Phys.\ Rev.\  D {\bf 16}, 1762 (1977).





\bibitem{Garriga:1994ut}
  J.~Garriga,
  Phys.\ Rev.\  D {\bf 49}, 5497 (1994)
  [arXiv:hep-th/9401020].

\bibitem{Dunne:2005rt}
  G.~V.~Dunne and H.~Min,
  Phys.\ Rev.\  D {\bf 72}, 125004 (2005)
  [arXiv:hep-th/0511156].


\bibitem{Volovik:2003fe}
  G.~E.~Volovik,
  Int.\ Ser.\ Monogr.\ Phys.\  {\bf 117}, 1 (2006).


\bibitem{Wong:2002rg}
  S.~M.~H.~Wong,
  hep-ph/0202250.








\bibitem{Skyrme:1961vq}
  T.~H.~R.~Skyrme,
  Proc.\ Roy.\ Soc.\ Lond.\ A {\bf 260}, 127 (1961).

\bibitem{Skyrme:1962vh}
  T.~H.~R.~Skyrme,
  Nucl.\ Phys.\  {\bf 31}, 556 (1962).


 \bibitem{Adkins:1983ya}
  G.~S.~Adkins, C.~R.~Nappi and E.~Witten,
  Nucl.\ Phys.\ B {\bf 228}, 552 (1983).


















\bibitem{Achucarro:1999it}
  A.~Achucarro and T.~Vachaspati,
  Phys.\ Rept.\  {\bf 327}, 347 (2000)
  [Phys.\ Rept.\  {\bf 327}, 427 (2000)]
  [arXiv:hep-ph/9904229].



\bibitem{Barriola:1993fy}
  M.~Barriola, T.~Vachaspati and M.~Bucher,
  Phys.\ Rev.\  D {\bf 50}, 2819 (1994)
  [arXiv:hep-th/9306120].
































\end{thebibliography}
\end{document}